# Information, Meaning, and Intellectual Organization in Networks of Inter-Human Communication [1]

Loet Leydesdorff
University of Amsterdam

## Introduction

Due to the salience of citations in bibliometrics, there have been periodic calls for a theory of citation (e.g., Amsterdamska & Leydesdorff, 1989; Cozzens, 1989; Cronin, 1981, 1984, 1998; Garfield, 1979; Kaplan, 1965; Leydesdorff, 1998; Leydesdorff & Amsterdamska, 1990; Luukkonen, 1997; Nicolaisen, 2007; Woolgar, 1991; Wouters, 1998, 1999). Theories about citations tend to emphasize the relational aspect—that is, citation *relations* among authors and/or documents. Relations can also be aggregated into networks and the citation networks can be analyzed using social network analysis (e.g., Hummon & Doreian, 1989; Otte & Rousseau, 2002). However, neither meaning nor knowledge is purely relational. Meaning, rather, is provided positionally, not relationally.

Unlike Shannon-type information—that is, the uncertainty in a probability distribution (Shannon, 1948, p. 10)—meaning can only be provided with reference to a system for which "the differences make a difference" (MacKay, 1969; Bateson, 1972, p. 315). I shall argue that systems can be considered as sets of relations that are the results of first-order relations. However, the sets relate at the systems level not in terms of individual relations, but in terms of *correlations*. Because of potentially spurious correlations among two distributions of relations given a third one, uncertainty can also be reduced in the case of interactions among three (or more) sources of variation (Strand & Leydesdorff, 2013; cf. Garner & McGill, 1956). This communication at the systems level can be expressed as mutual information in the overlap among the sets—or with the opposite sign as reduction of uncertainty because of mutual redundancies.

---

On top of the information and meaning exchanges, discursive knowledge develops by relating meanings reflexively on the basis of cognitive codes that remain mentally and socially constructed (Callon *et al*., 1986). The specification of the role of citations in the development of discursive knowledge thus first requires that the relational perspective be extended with a positional one. Positions make it possible to develop perspectives (Leydesdorff & Ahrweiler, 2014). Translations among perspectives provide a third layer of the exchange on top of information processing in relations and the redundancy generated when meanings are shared.

**Meaning, Meaningful Information, and the Codification of Meaning**

One can provide the Shannon-type information contained in relations with a variety of meanings from different perspectives. A perspective, however, presumes a position. In the case of a reflecting agent, each position is defined in terms of the vector space that is spanned—as an architecture—by the set(s) of relations (Leydesdorff, 2014a). When a distributed network reflects (e.g., discursively), the positioning contains uncertainty since different (and potentially orthogonal) perspectives can be used at the same time, but from different positions. The meaning of the information for the receiving system can then no longer be identified unambiguously, but can only be hypothesized with reference to a virtual domain of possible relations and meanings. Giddens (1979, p. 64) called this virtual structure "an absent set of differences". The latent dimensions can be considered as providing perspectives that allow for sharing or not-sharing meaning(s) when information is positioned in a network.

For example, a perspective can be used to develop discursively a rationalized system of expectations, and thus to generate knowledge by codifying specific meanings. The codification provides an additional selection mechanism: perspectives thus add a third layer by potentially codifying communication on top of the information and meaning processing. In this context, the notion of "double contingency" (Parsons, 1968, p. 436; Parsons & Shills, 1951, p. 16) can be extended to a "triple contingency" (Strydom, 1999, p. 12). Meaningful information can first be selected from the Shannon-type information fluxes on the basis of codes that are further developed in the communications. The three layers operate in parallel.

The construction of this triple-layered system is bottom-up, but—using a cybernetic principle—control can increasingly be top-down as the feedback layers are further developed (Ashby, 1958). Whereas the three contingencies can be expected to develop in parallel, this assumption enables us to hypothesize a hierarchy among the layers that can be expected for analytical reasons. Let me stepwise extend the single-layered and linear Shannon-model (Figure 1.1 below) into such a triple-layered model, as depicted in Figure 1.2.

## Extensions of the Shannon-Weaver Model

As is well known, Shannon (1948, p. 3) first focused on information that was not (yet) meaningful: "Frequently the messages have *meaning*; that is they refer to or are correlated to some system with certain physical or conceptual entities." According to Shannon (1948, p. 3), however, "(t)hese semantic aspects of communication are irrelevant to the engineering problem."

It is less well known that Shannon's co-author Warren Weaver argued that Shannon's distinction between information and meaning "has so penetratingly cleared the air that one is now, for the first time, ready for a real theory of meaning" (Shannon & Weaver, 1949, p. 27). Weaver (1949, p. 26) proposed to insert thereto another box with the label "semantic noise" into the Shannon model between the information source and the transmitter, as follows (Figure 1.1):

[Figure 1.1 here]

What if one adds a similar box to the right side of this figure between the receiver and the destination of the message (added in grey to Figure 1.1)? The two sources of semantic noise may be correlated; for example, when the sender and receiver of the message share a language or, more generally, a code of communication. I propose to distinguish between "language" as the natural—that is, undifferentiated—code of communication versus codes of communication which can be symbolically generalized and then no longer require the use of language (Luhmann, 2002 and 2012, pp. 120 ff.). For example, instead of negotiating about the price of a commodity, one can simply pay the market price using money as a symbolically generalized medium of communication. One is able to translate reflexively among codes of communication by elaborating upon the different meanings of the information in language (Bernstein, 1971).[2]

Thus, one arrives at the following model (Figure 1.2):

[Figure 1.2 here]

[2] I deviate here from Luhmann's theory. In his theory, the sub-systems of communication are operationally closed and communications cannot be transmitted reflexively from one system into another (cf. Callon, 1998; Leydesdorff, 2006 and 2010a).

Contrary to Shannon's counterintuitive definition of information as uncertainty, MacKay (1969) proposed to define information as "a distinction which makes a difference," and Bateson (1972, p. 315) followed by defining information as "a difference which makes a difference" to which he added "for a later event" (p. 381). In my opinion, a difference can only make a difference for a system of reference receiving the information. This latter system may be able to provide a relevant part of the Shannon-type information with meaning from the perspective of hindsight—that is, at a later moment. Meaningful (Bateson-type) information can no longer be considered as Shannon-type information, since it is a selection from the uncertainty that prevails. Bateson-type information may add to the uncertainty, but it can also be "informative" and thus reduce uncertainty for the receiving system (Brillouin, 1962).

In other words, one can distinguish between "meaningful information"—potentially reducing uncertainty—and Shannon-type information that is by definition equal to uncertainty (Hayles, 1990, p. 59). Shannon (1948) chose his formulas so that uncertainty could be measured as probabilistic entropy in bits of information. The mathematical theory of communication thus provides us with entropy statistics that can be used in different domains (Bar-Hillel, 1955; Krippendorff, 1986; Theil, 1972). Meaning is provided to the information from the perspective of hindsight (of the "later event"—that is, a system of reference). However, the measurement of "meaningful information" in bits or otherwise had remained hitherto without an operationalization (cf. Dretske, 1981).

## The Cybernetic Perspective

The semantic noises can be correlated when the semantics are shared, for example, in a common language. Various forms of semiotics have been developed to study the processing of signs in inter-human communication (e.g., Fiske, 2011, pp. 37-60; Nöth, 2014). The focus of this contribution, however, remains on the shaping of discursive knowledge using cybernetic and information-theoretical perspectives. Can the effects of the codification in scholarly exchanges also be measured?

The sharing of meaning is far from error-free, and thus other uncertainty can be generated at this later moment, but the selective operation is analytically different from the generation of variation: some differences are selected as making a difference—a signal—whereas other differences (bits) are discarded as noise. A second contingency is thus added reflexively to the relational uncertainty in the communication of information.

The relations are “contingent”—and not necessary or transcendent—because a variation could also have been different. Secondly, the relational information may mean something different for the sender and the receiver, but this is again contingent because it depends on the respective positions in the networks of relations. However, both analysts and participants are able to specify an *expectation* about this meaning, given codes of communication, insofar as the codes have emerged as densities (eigenvectors) in the networks of communications at the two lower levels of relational information processing and positional meaning-sharing (Leydesdorff, 1998).

Parsons (1951, p. 10f.) elaborated “double contingency” as a basic condition for inter-human interactions, but he presumed a normative—that is, relatively stable—binding of mutual expectations in a symbolic order (Deacon, 1997). However, different horizons of meaning can always be invoked (Husserl, 1962; Luhmann, 1990, p. 27 and 1995, p. 69). This third layer of codes in the communication emerges as a source of friction—and thus contingency—when differences become manifest in historical encounters such as misunderstandings. Normative integration is then no longer sufficient, and differentiation among the codes of communication can become functional. For example, while concerned about “truth,” science is not involved in the pursuit of religious truth. “Truth-finding” in a criminal investigation is differently coded from heuristics in theoretical contexts.

The symbolic order among the codes of communication is not a given, but a construct that can be reconstructed reflexively by using another code such as another alphabet or language. Luhmann (1995) added that symbolically generalized codes of communication can be functionally different. Whereas normative integration was presumed in understanding at the second level—using a common language—differentiation operates against the integrative tendency of normative learning by developing cognitive learning in parallel. When this differentiation prevails, the fluxes of communication can no longer be integrated historically into organizations, but tend to “self-organize.”

Against Luhmann’s reification of these tendencies (Habermas, 1987; Leydesdorff, 2006 and 2010a), I propose to consider the self-organizing dynamic as a third contingency (Strydom, 1999): a triple contingency can thus be expected to operate in inter-human communications, but the processing at different levels remains historically contingent since socially constructed. The self-organizing tendencies have the status of hypotheses; the codes can be expected to enable both participants and analysts to specify expectations (Leydesdorff, 2012).

In other words, inter-human communication first requires a historical medium in which probabilistic entropy (Shannon-type information) is generated, but this first-order proliferation of differences can be provided with meaning at both the sending and receiving ends. Meaning can be provided to the communication from the perspective of hindsight, but also differently using other perspectives and codes with reference to self-organizing "horizons of meaning" (Husserl, 1962; Luhmann, 1995, pp. 60 ff.; cf. Borch, 2011, p. 41).

Note that the codes of communication can be considered as second-order variables, that is, variables that are attributed as eigenvectors to the communications as first-order variables (Von Foerster, 2003).[3] Consequently, the coded dimensions of the communication can no longer be attributed to the communicating agents; they are attributes of the communications and the analysis thus becomes more abstract and layered: not only can the agents interact, but also their interactions can be expected to interact. The next-order interactions among interactions provide the lower-level structures of first-order interactions with new degrees of freedom in feedback loops.

In summary, this model—based on and inspired by Luhmann (1995)—follows Herbert Simon's (1973) model of complex systems, but with modifications. One assumes both horizontal and vertical differentiation in the communication. Vertical differentiation was visualized in Figure 1.2 and can be labeled as (i) *interactions* at the bottom providing variation, (2) *organization* of the communication when the different codes of communication are historically interfaced, and (3) *self-organization* of the codes of communication spanning horizons of meaning (Luhmann, 1975).

Horizontally, the codes of communication can be expected to operate in parallel; they can be considered as the evolving units and are modeled as "genotypical." Because the codes are not material ("phenotypical"), they can develop with a higher frequency than the historical realizations. Expectations proliferate faster than actions (Weinstein & Platt, 1969). In this respect, the model is different from Simon's model where the higher the level, the lower the frequencies. The additional feed forward of the communication under the condition of horizontal differentiation among the codes enables the communication to process more complexity. When the normative order among the codes is broken, differentiation can evolve into another degree of freedom in the system's capacity (Leydesdorff, 2014b).

[3] Luhmann indicates the latent dimensions with the word "eigenvalue". Technically, the eigenvalue of an eigenvector is the factor by which the eigenvector is scaled when multiplied by the matrix.

The uncertainty can be reduced by the specification of expectations in highly codified communications such as systems of rationalized expectations or, in other words, scholarly discourses. Translations from one code into another require integration into elaborate discourse in a historical context (at the second level), but not necessarily at the same moment. The historical organization can thus be considered as a synchronizing retention mechanism of the otherwise self-organizing dynamics. While these mechanisms can be distinguished analytically, they operate in parallel and can be expected to "overflow" (Callon, 1998) into one another because of the ongoing generation of uncertainty in all historical processes.

## Relevance for the Study of Organized Knowledge Production in the Sciences

The distinction between organization and self-organization of communication enables us to operationalize distinctions that were made in science studies, but could at the time not yet be operationalized in communication-theoretical terms. In the sociology of science, for example, Whitley (1984) distinguished between the social and intellectual organization of the sciences or, in other words, between the "field"-level and the "group"-level (Rip, 1981). In the philosophy of science, Popper ([1935] 1959) introduced the distinction between the locally contingent context of discovery and the trans-local context of justification (cf. Lakatos & Musgrave, 1970). The field-level, the intellectual organization, and the context of justification are evolutionary and self-organizing (Popper, 1972); whereas the group-level, the social organization, or the contexts of discovery are historically organized. The two levels co-evolve and are co-constructed, but the direction of the arrows is reversed (Campbell, 1960).

For example, when the peer-review process is organized in terms of editors and referees at the journal level, this is a social process, but the intellectual organization is supposed to take control in terms of the codes of the communication. The codes of communication are needed for the context of justification in order to function, but the material conditions also need to be organized. The social organization of science is sensitive to funding, but the intellectual organization in terms of self-organizing codes of communication can be expected to resist such steering of the scientific enterprise (van den Daele & Weingart, 1975). The intellectual self-organization operates as a latent feedback mechanism. Under certain conditions, this feedback can come to fruition into a feed-forward, and the field can auto-catalytically develop its code(s) of communication (Figure 1.3).

[Figure 1.3 here]

Figure 1.3 elaborates on Ulanowicz's (2009, p. 1888) model of auto-catalysis (cf. Padgett & Powell, 2012): a third code—that is, meaning providing system or perspective—can auto-catalyze the relation between the other two. However, the rotation can be clockwise or counter-clockwise (Ivanova & Leydesdorff, 2014, p. 930). Whereas the one dynamic can be appreciated as a feed-forward from organization at each moment of time to self-organization over time, the reverse dynamic retains historical organization at each moment of time. Since both dynamics can be expected to operate in parallel but opposite directions, one can assume a balance or trade-off between them: is intellectual self-organization leading at the field-level or historical organization at the institutional level? Note that one can only observe the historical instantiations; the self-organization remains a theoretically-informed hypothesis about an evolutionary (that is, supra-historical) dynamic.

In another context—that of the Triple Helix of university-industry-government relations—I proposed mutual information in three (or more) dimensions as an indicator of this trade-off between historical organization (in networked university-industry-government relations) versus the evolutionary self-organization of synergy in terms of functionalities such as—in the case of Triple-Helix relations—(*i*) novelty production through the development of science and technology, (*ii*) economic wealth generation, and (*iii*) normative control by governance (Leydesdorff & Zawdie, 2010; cf. Ulanowicz, 1986, p. 143). The historical relations cannot be the sole purpose of a Triple Helix, but one rather aims at the fruition of these relations into synergy at a systems level. Under what historical conditions can the loops among the three juxtaposed coordination mechanisms flourish and blossom auto-catalytically?

$$T_{123} = H_1 + H_2 + H_3 - H_{12} - H_{13} - H_{23} + H_{123} \quad (1)$$

Mutual information in three dimensions (Eq. 1) can be used to model the trade-off between organization and self-organization because this measure can be positive or negative. The equation can be derived from the Shannon formulas (e.g., Abramson, 1963; McGill, 1954;

cf. Jakulin, 2005; Yeung, 2008), but $T_{123}$ can no longer be considered as a Shannon entropy because it can also be negative (Krippendorff, 2009a). Shannon's model (Figure 1.1), however, excluded feedback loops and thus developments against the arrow of time—in accordance with Shannon's aim to discard meaning-processing as not relevant to the engineering problem.

Leydesdorff and Ivanova (2014) showed that the mutual information in three (or more) dimensions can also be considered as a measure of mutual redundancy—that is, overlap among "pure sets" (*ibid*., p. 391). An overlap among sets is then appreciated twice (or more times) by considering both overlapping systems as systems of reference. It could then be shown that the mutual redundancy $R_{12} = - T_{12}$ in the case of two systems, while in the case of three systems $R_{123} = T_{123}$ (with the opposite sign). The choice of sign warrants consistency with Shannon's (1948) mathematical theory of communication, so that the values can be expressed in bits of information (Leydesdorff, 2010b). Negative values of $R$ indicate reduction of uncertainty because of synergy in the configuration of relations.

Given space constraints, I will not repeat this argument, but instead use the mutual redundancy in three dimensions as a possible operationalization for the distinction between (self-organizing and hypothesized) intellectual versus historical organization in texts using, on the special occasion of this *Festschrift*, the work of Professor Blaise Cronin. This œuvre provides an example of a historically organized set of documents in which intellectual organization operates reflexively to the extent that it can be expected to prevail over the historical organization of the texts.

To what extent are these documents organized intellectually in terms of title words, cited references, and/or the title words of the papers citing them? Can one use the concepts of latent variables (factors or eigenvectors) of the matrices of documents versus words to uncover this trade-off between intellectual self-organization over time and social or semantic organization at specific moments of time? I operationalize the three layers specified above as follows: (1) relations in terms of co-occurrences of title words, (2) the positions of these words in the vector-space spanned by these relations, and (3) the mutual redundancies among the three main (factor-analytic) dimensions of this vector-space in each set.

## Data

Given the character of this *Festschrift* for Professor Cronin, it seemed reasonable to illustrate the above arguments empirically by focusing on this author's œuvre insofar as available using the Web of Science (WoS) data provided by Thomson-Reuters. Since there are several authors under "B Cronin" in WoS, the download was limited to "au = Cronin B* and ci=Bloomington". Cronin has published at this address since 1991. Thus, 164 documents were retrieved from the database on April 23, 2014. I use this data and the 949 articles citing these 164 documents at this same date. Table 1.1 provides descriptive statistics.

[Table 1.1 here]

The sets of documents are used as samples to pursue an analysis analogously to the evaluation of aggregated journal-journal citations (Leydesdorff, 2011a) and for title words in a single journal—namely, *Social Science Information* (Leydesdorff, 2011b).

## Methods

Three matrices are central to the analysis:

1. The asymmetrical word/document matrix based on the 164 documents authored by Cronin as cases, and the 57 title words in these documents that occurred more than twice in this set (after correction for stopwords, using a list of 429 stopwords)[4];
2. The asymmetrical word/document matrix based on the 949 documents citing one of these 164 documents (1,441 times) versus the 108 words that occurred more than ten times in the titles of these citing documents (after a similar correction for stopwords); and,
3. Parsing the 3,526 cited references in the first document set[5], 398 cited source names could be retrieved, of which 109 (27.4%) matched with the abbreviations for journal names used in the *Journal Citation Index* 2012 of WoS.[6] These 109 journal names were used as variables to the 164 documents as cases for the construction of a third matrix.

---

[4] Provided at http://www.lextek.com/manuals/onix/stopwords1.html

[5] Table 1.1 provides the number of 3,503 cited references based on cross-tabulation in Excel, and using the field "N of references" (NRef) in the WoS output.

[6] Using automatic matching, the *Journal of the American Society for Information Science* (*JASIS*) is not matched because it is included in JCR 2012 as the *Journal of the American Society for Information Science and Technology* (*JASIST*). However, 163 references in the set refer to this old title. We will use this set as an additional control in the discussion section.

Co-occurrence matrices and cosine matrices were derived from each of these three matrices for further analysis and visualization using Pajek (v. 3.11).[7] The three matrices can be used for drawing semantic maps, both in terms of relations and in terms of cosine-normalized relations in the vector space.[8] Moreover, the asymmetrical word/document matrices can be imported in SPSS (v. 21) for factor analysis. Factor loadings on the three main components (after orthogonal rotation using Varimax) are used for visualizing the variables (vectors) in relation to the first three eigenvectors and also for the analysis of mutual redundancy using dedicated software.

## Results

*The Document Set Authored by Cronin* (*N* = 164)

As noted, 164 documents were downloaded on April 23, 2014, using the search string "au = Cronin B* and ci= Bloomington". These documents contain 57 title words which occur more than twice after correction for stopwords. Figure 1.4 shows the *relational* network among 56 of these words colored according to the partitioning using Blondel *et al.*'s (2008) algorithm for community-finding, and Kamada and Kawai's (1989) algorithm for the layout.

[Figure 1.4 here]

A relational map of co-occurring words in the same subject area can always be provided with an interpretation because the words are grouped and placed in relation to one another. Frequently used words will tend to be central (e.g., "Science," "Society," "Library"). In this set, for example, "Bibliometrics" is placed in this central set, but in a grouping differently from words which are commonly used in bibliometrics such as "Author," "Journal," and "Citation."

After cosine-normalization and setting a threshold of cosine > 0.2, one obtains a systems perspective on Cronin's œuvre. Fifty-three of the title words form a largest component (Figure 1.5); five communities are indicated using the algorithm of Blondel *et al*. (2008)[9] with a modularity $Q = 0.542$. The modularity of this network is enhanced because of the threshold; the words are now grouped in the vector space (Leydesdorff, 2014a). The grouping indicates the structure in the set of relations.

---

[7] Pajek is a program for network analysis and visualization; available for download at http://pajek.imfm.si/doku.php?id=download .

[8] The cosine can be considered as the non-parametric equivalent of the Pearson correlation; as against the latter, the distribution is not first z-normalized with reference to the mean (Ahlgren *et al*., 2003).

[9] This algorithm is used because the algorithm of VOSviewer indicated three more communities.

Thus, we have moved from a relational to a positional perspective on the structure in this data (Burt, 1982; Leydesdorff, 2014a). The topology is different: we no longer study the network of *relations* among words in terms of co-occurrences ("co-words"; Callon *et al*., 1983), but the *correlations* among the distributions of words over the documents under study. The grouping of words in Figure 1.5 indicates the latent dimensions of the network as a system of words (Leydesdorff, 2014a).

[Figure 1.5 here]

In Figure 1.5, for example, "Bibliometrics," "Library Studies," "Education," and "Management" are grouped (using pink) as different from bibliometric terminology such as "Citation," "Analysis," "Measure," "Author," and "Journal." The differences between Figures 1.4 and 1.5, however, are in this case not so large.

Figure 1.6 uses a different input: it visualizes the three-factor matrix based on the same set (Vlieger & Leydesdorff, 2011). For reasons of presentation, I have removed the negative (dotted) lines from the visualization and also the nine words which thus became isolates. All 57 words and their factor loadings were used in the further analysis of the mutual redundancy in three dimensions. Whereas these dimensions could be induced from Figure 1.5, I now force the three (latent) dimensions to become center stage. As noted, the choice for three is made for reasons of parsimony, but one can also extend this to more than three dimensions.

[Figure 1.6]

Factor 1 groups the words in bibliometrics; factor 2 the words focusing on scholarly communication; and factor 3 more general terminology. The three factors can be considered as the latent dimensions (eigenvectors) of the word/document matrix.

Eq. 1 can be used for the computation of the mutual redundancy among these three dimensions (Leydesdorff, 2010b). The (binned) factor loadings of the 57 words as variables provide a mutual redundancy of –1888.9 mbits of information. In other words, the uncertainty in this textual domain is reduced by almost two bits by the intellectual organization of the words in the three main (latent) dimensions.

*Citing Papers* (*N* = 949)

Using the 949 documents that could be retrieved as citing at least one of the 164 documents authored by Cronin, a similar procedure was followed. Figure 1.7 shows 92 of the 108 words occurring more than ten times in these documents with at least one (among three) *positive* factor loadings, similarly to Figure 1.6.

[Figure 1.7]

Figure 1.7 shows the structure in the vocabulary of Cronin's (citing) audiences. Bibliometric terminology loads on a second factor after a first one with a focus on academia; factor 3 indicates concerns of library and education.

Following an analogous procedure, the mutual redundancy among the three main dimensions in this matrix of 949 documents versus 108 title words is –70.1 mbits of information. This is only 3.7% of the synergy retrieved from the word distributions in the 164 cited documents that were authored by Cronin himself.

*Cited References*

The document set of 164 documents authored by Cronin is not only cited, but also citing. As noted, the documents contain 3,526 cited references. Since the cited references in WoS do not contain title words, I used the subfield of the abbreviated journal titles in the references as variables to the 164 documents. This can be considered as a representation of the knowledge bases of Cronin's articles (Leydesdorff & Goldstone, 2014).

Among the 3,526 cited references, 398 unique sources can be counted,[10] of which 109 sources could be matched to the journal abbreviations provided by the Journal Citation Reports for 2012. One can thus construct two matrices: one with 398 cited sources as variables and another with 109 matched sources that occur in 1,223 (34.3%) of the cited references.

[Figure 1.8 here]

Figure 1.8 shows the map of the factor matrix of 96 of these 109 journals based on 91 documents carrying these references.[11] Factor 2 is recognizable as a group of information-science journals, but the designation of the other two factors is less obvious except that Factor 1 includes general-science journals such as *Nature, The Lancet,* and the *American Economic*

[10] A referenced journal has to be included more than once into the set so that two journals are related by the document as the observational unit.

[11] The 3,526 cited references to 398 sources were counted in 111 documents.

*Review,* whereas Factor 3 is composed mainly of specialist journals in the social sciences and the humanities.

[Table 1.2 here]

The mutual redundancies are +14.2 mbit for the larger set of 398 cited sources versus -160.4 mbit for the references to journals active in the WoS database. Thus, these more codified references contribute to the synergy, while the larger set tends to be more incidental and contingently organized. Table 1.2 summarizes the findings for the four analyses discussed above.

## Discussion

As noted, I performed a similar analysis in a contribution on the occasion of the 50$^{th}$ volume and publication year of *Social Science Information* (SSI) using title words in the volumes between 2005 and 2009 (Leydesdorff, 2011b). Using 69 title words occurring three or more times in a set of 149 titles, the mutual redundancy among the three main dimensions of this matrix added +50.6 mbits to the uncertainty. These 149 documents were cited by 187 other documents; for the title words in these citing journals I obtained a mutual redundancy of –106.2 mbits. In this case, the citing journals provide windows on different (self-organizing) literatures, whereas the articles published in the journal were intellectually a heterogeneous set that was organized historically.

In the case of Cronin's publications, the original documents are all authored by him and thus the title words are intellectually organized to a degree much larger than the citing documents. The latter show a synergy comparable to that of the set of citing papers in the case of *SSI* (–70.1 versus –106.2 mbits). When the non-source references are included in the analysis of Cronin's set, the synergy disappears, while it remains when the analysis is restricted to the set of references to indexed journals.

We note another possible control: Adding the abbreviation "*J AM SOC INFORM SCI*", that is, the name of *JASIST* before 2001 (but no longer included in JCR and therefore not matched above), another 163 references can be included (1223 + 163), and the journal set is extended to (109 + 1 =) 110. The mutual redundancy is in this case further increased to –179.5 mbits. This result accords with the expectation that references to *JASIS* contribute to the intellectual organization of the set.

## Conclusions

Luhmann's sociological theory of communication and Shannon's mathematical one can be considered two almost orthogonal perspectives. On the one hand, Luhmann (1995, p. 67) defined information as a selective operation and stated that "all information has meaning." Thus, the measurement of communication (e.g., in bits of information) remains external to this theoretical perspective. On the other hand, Shannon (1948, p. 3) excluded "meaning" as not relevant to his theory of communication. The crucial question, in my opinion, is how meaning is generated in communication of information and then also codified. Can the one perspective be translated into the other or are these theories fully incommensurable?

Weaver's (1949, p. 27) call for a "real theory of meaning" based on Shannon's distinction between meaning and information can be elaborated both theoretically and then also empirically. We have begun to develop instruments such as semantic maps for the positioning of information, and mutual redundancy for the measurement of the relations among codes in the communication. These operationalizations have been illustrated empirically.

The first step in how meaning is generated in communicative relations is articulated in the operation of *semantic* mapping. The aggregate of relations allows for a systems perspective since an architecture is shaped by the network which can also be analyzed in terms of correlations and latent dimensions. The relational analysis can thus be complemented with a positional one (see Leydesdorff, 2014a). Meaning is provided in terms of positions, that is, with reference to a system. The system(s) of reference position the incoming information and thus appreciate uncertainty as noise or signal. Over time, this positioning may either increase or decrease uncertainty within the system. Brillouin (1962) introduced the concept of "negentropy" in this context.

Negative entropy can be generated when the redundancy increases more rapidly than uncertainty, given that the maximum entropy—that is the sum of the redundancy and uncertainty—can also evolve in dynamic systems (Brooks & Wiley, 1986, p. 43). As Krippendorff (2009b, p. 676) formulated: "Note that *interactions with loops entail positive or negative redundancies, those without loops do not*. Loops can be complex, especially in systems with many variables."

Using Weaver's (1949, p. 26) loophole of "semantic noise", next-order loops can be related to the Shannon model. Using the sociological progression from Parsons' assumption of normative integration in the first next-order loop to Luhmann's option of functional differentiation in a second-order loop of codes of communication, a model with both horizontal and vertical differentiation (Luhmann's social or systems differentiations, respectively) could thus be developed in terms that allow for empirical operationalization.

The codes of communication provide a superstructure that operates evolutionarily (as "genotypes"), and thus becomes historically manifest only as a source of structural reduction of uncertainty (i.e., redundancy). When different codes of communication operate, the same information may redundantly be provided with different meanings and thus appreciated twice or more times. In the case of three or more codes, two rotations are possible (Figure 1.3 above), of which one can be considered as feed forward and the other as feedback. Using the example of Cronin's publications, we have suggested that mutual redundancy can be used as a measure of intellectual versus historical (in this case, textual) organization.

Unlike the organization of articles in journal issues, the single author adds intellectual organization to his texts. The titles are not a bag of words which can be co-occurring or not, but their organization can be made visible as meaningful using a semantic map, and then further be analyzed in terms of the synergy among the latent dimensions of the vector space spanned by the distributions of words as variables in relation to their textual organization—that is, with the historical documents as the cases (Hesse, 1980, p. 103; Law & Lodge, 1984; Leydesdorff, 1997). The author organizes this vector space intellectually by more than an order of magnitude when compared with the cited references or the documents that cite his œuvre.

**Acknowledgement**

I thank Cassidy Sugimoto, Lutz Bornmann, and an anonymous referee for comments on a previous version.

**Tables and Figures**

Table 1.1. Descriptive statistics of the downloads under study, including the number of cited and citing documents.

| | *N* | Times Cited | Cited References |
|---|---|---|---|
| Article | 65 | 1113 | 2314 |
| Article; Proceedings Paper | 7 | 151 | 187 |
| Biographical Item | 2 | 9 | 2 |
| Book Review | 36 | 1 | 78 |
| Discussion | 1 | 0 | 0 |
| Editorial Material | 35 | 36 | 51 |
| Letter | 7 | 17 | 16 |
| Meeting Abstract | 1 | 0 | 0 |
| Note | 4 | 72 | 75 |
| Review | 6 | 42 | 780 |
| | **164** | **1441**[12] | **3503** |

Table 1.2. Mutual redundancy among the three main dimensions of the four document/word matrices compared (in mbits of information).

| | *Mutual redundancy in mbits* |
|---|---|
| *164 documents authored by Blaise Cronin* | – 1,888.9 |
| *949 citing documents* | – 70.1 |
| *398 cited sources* | + 14.2 |
| *109 cited sources that match with JCR* | – 160.4 |

[12] These 1441 citations—based on aggregating the field "times cited" of the 164 documents—were carried by 949 citing documents (including self-citations).

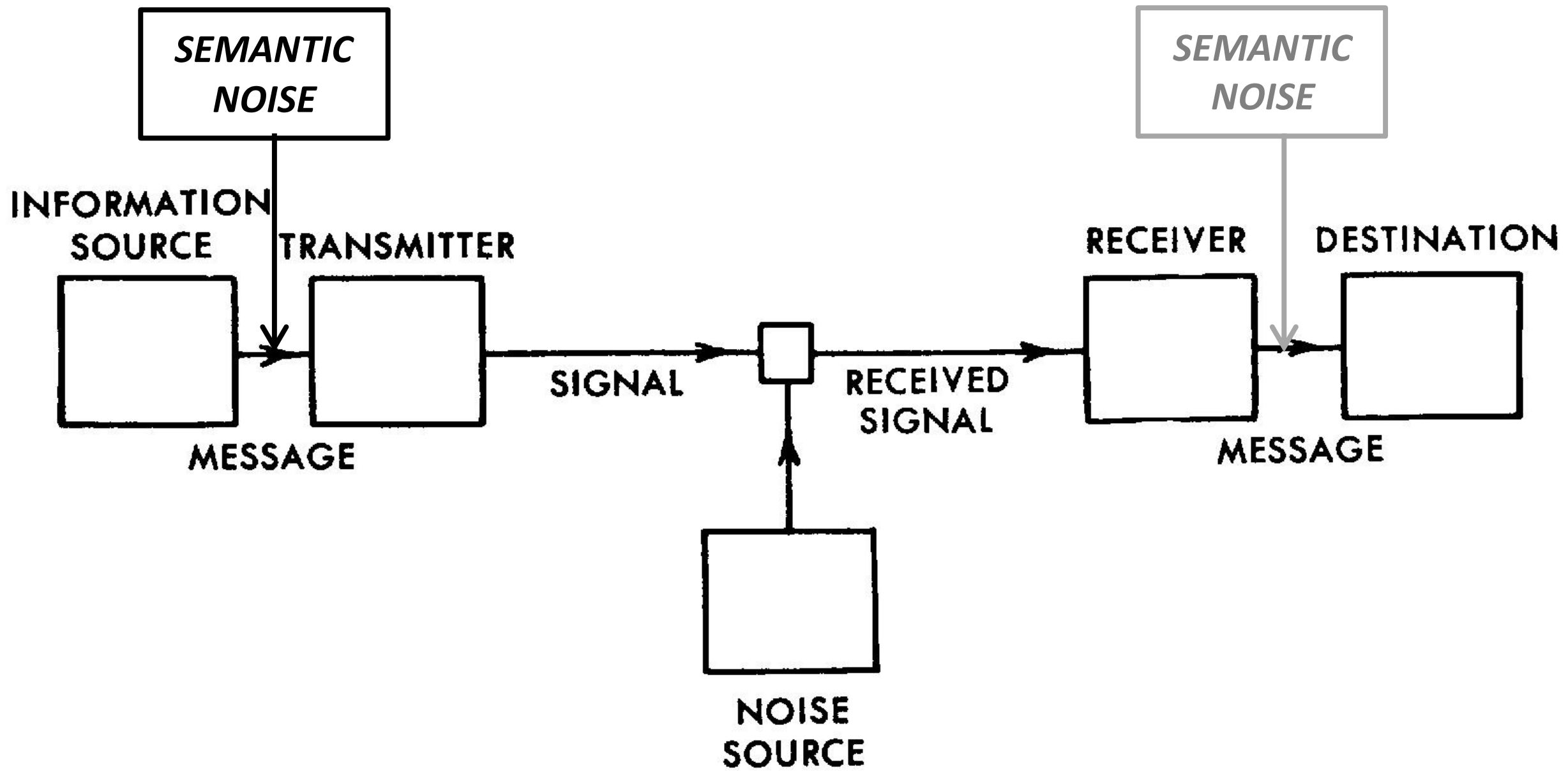

Figure 1.1. Schematic diagram of a general communication system. Source: Shannon (1948, p. 380); with Weaver's box of "semantic noise" first added (to the left) and then further extended with a second source of "semantic noise" between the receiver and the destination (to the right).

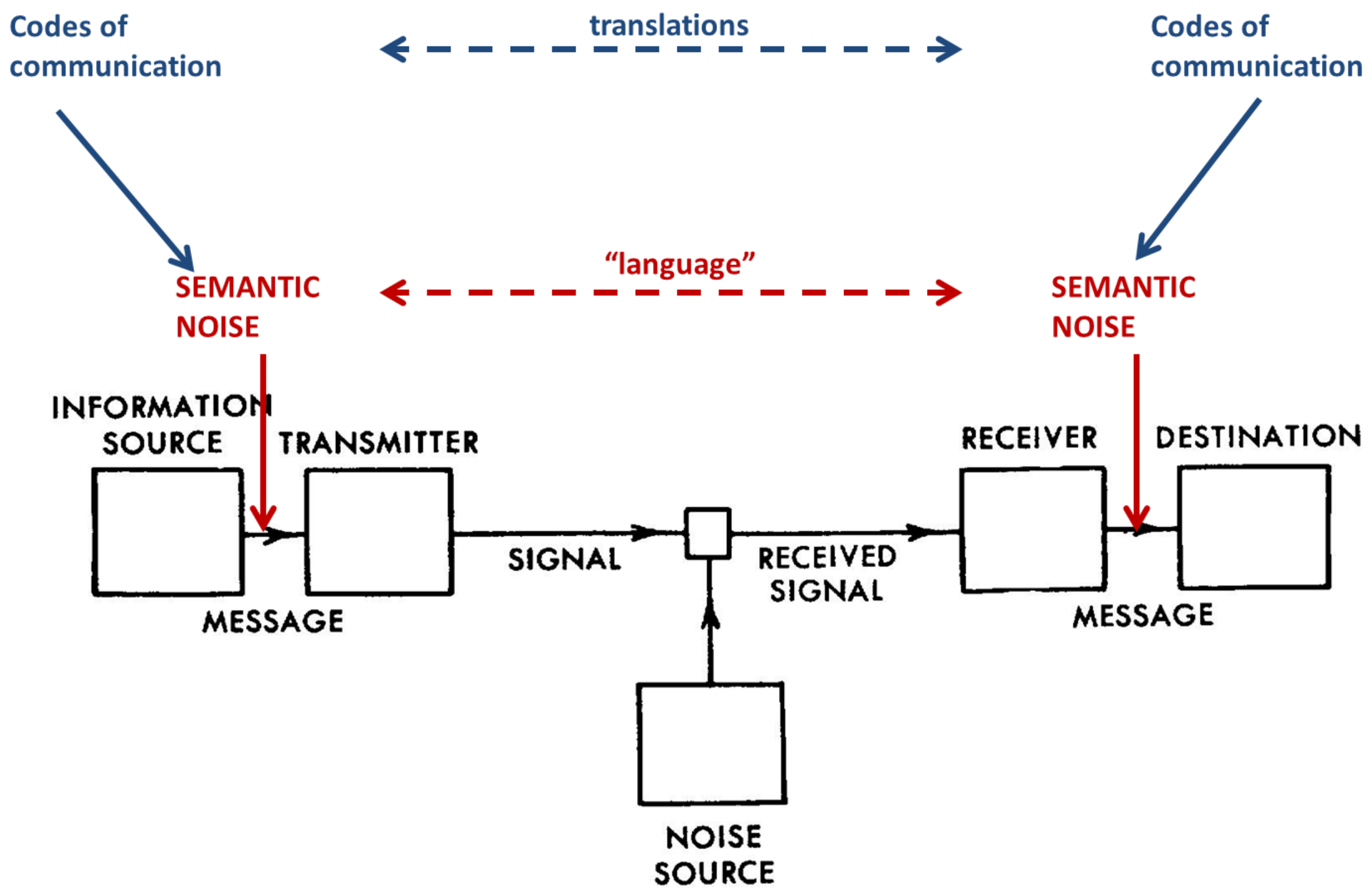

Figure 1.2. Three mutual contingencies in the dynamics of codified knowledge.

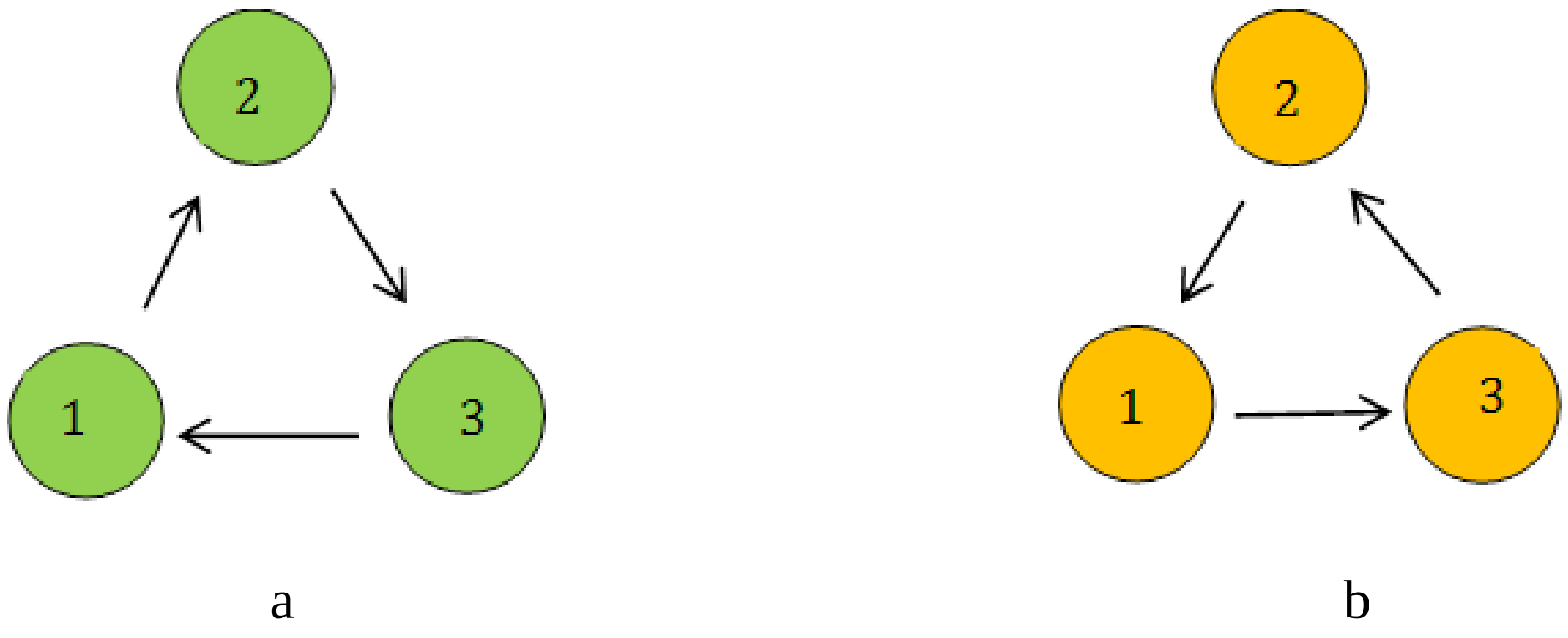

Figure 1.3. Circulation and feedback in cycles in both directions.

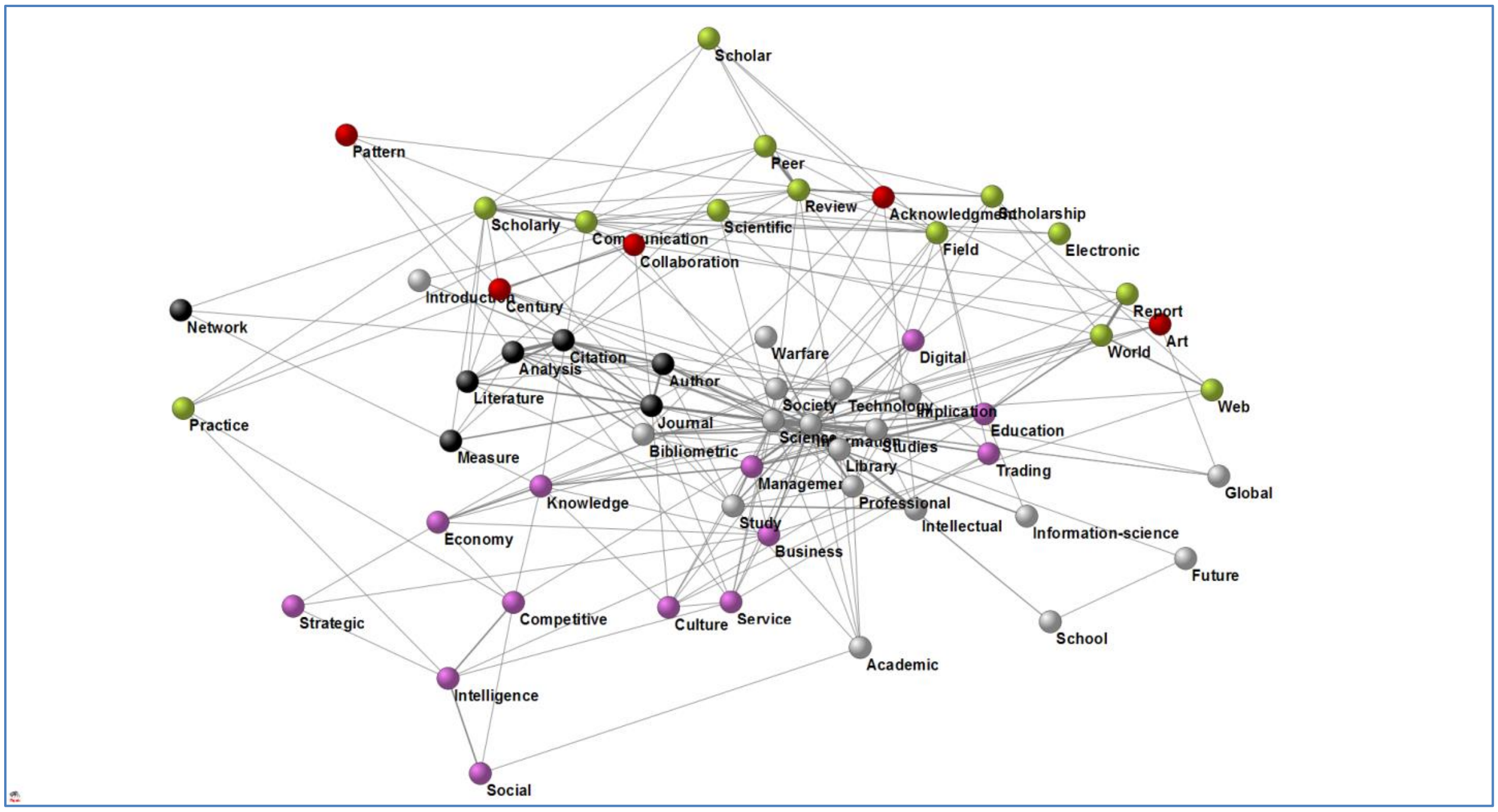

Figure 1.4. 56 (of the 57) words connected in the largest component of the network of title-words occurring more than twice in the set. *Q* = .359; *N* of Clusters = 5.

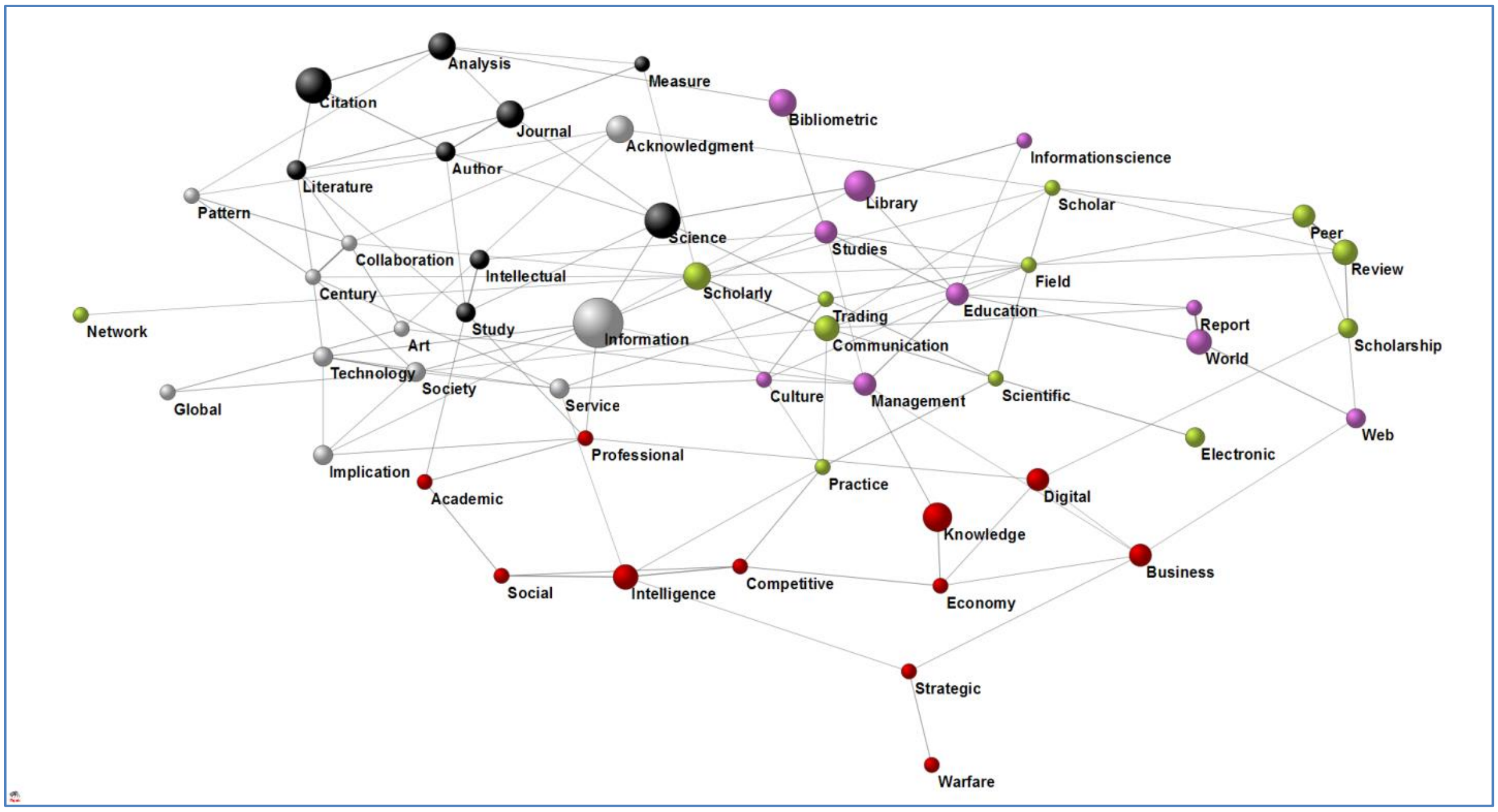

Figure 1.5. 53 words organized in five communities forming a largest component using 164 documents (cosine > 0.2).

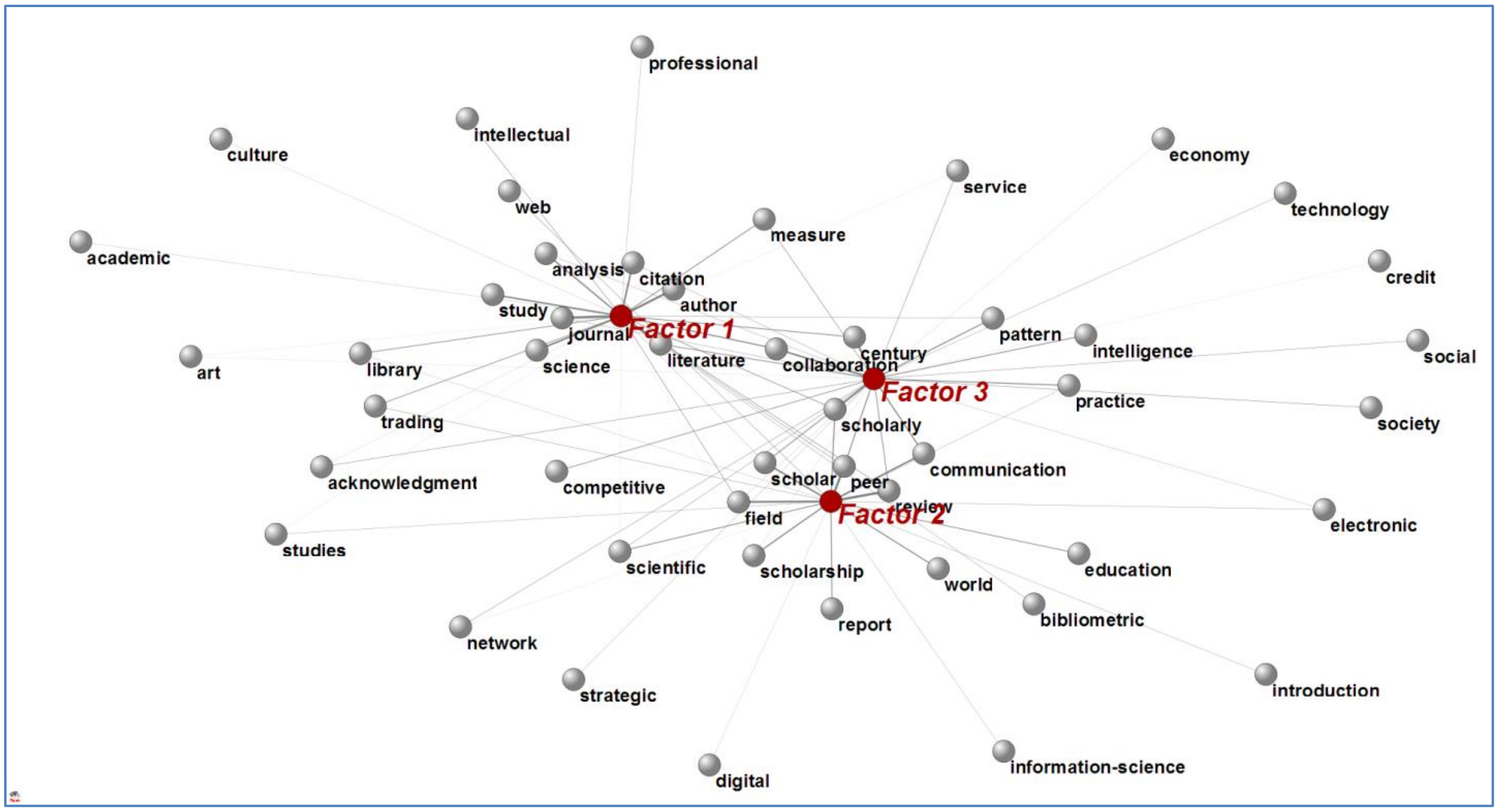

Figure 1.6. 48 words with positive loadings on three factors in a matrix of 164 documents and 57 variables (words).

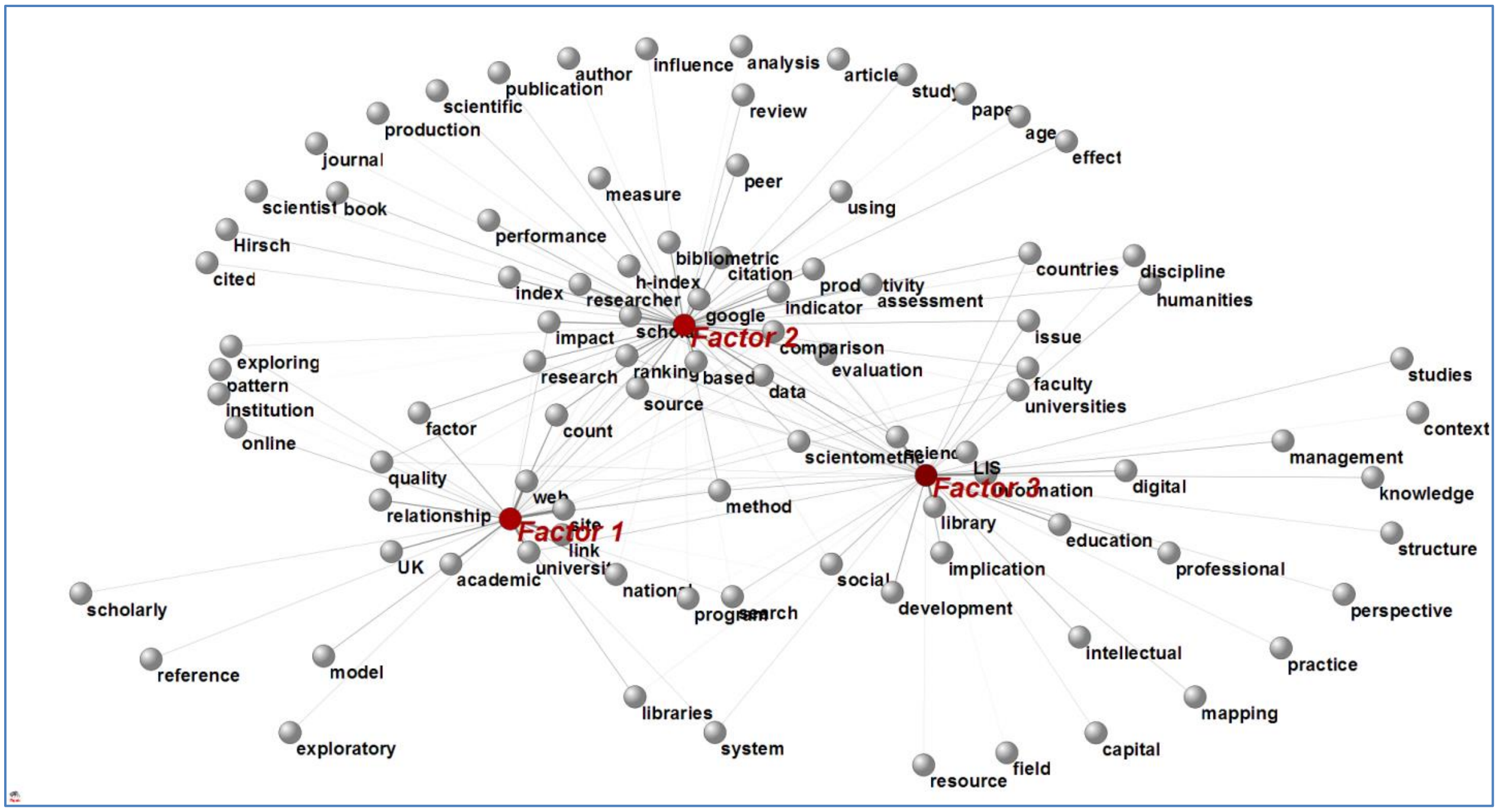

Figure 1.7. 92 (of the 108) words occurring more than ten times in the 949 citing documents in relation and positive factor loading on at least one of the three factors.

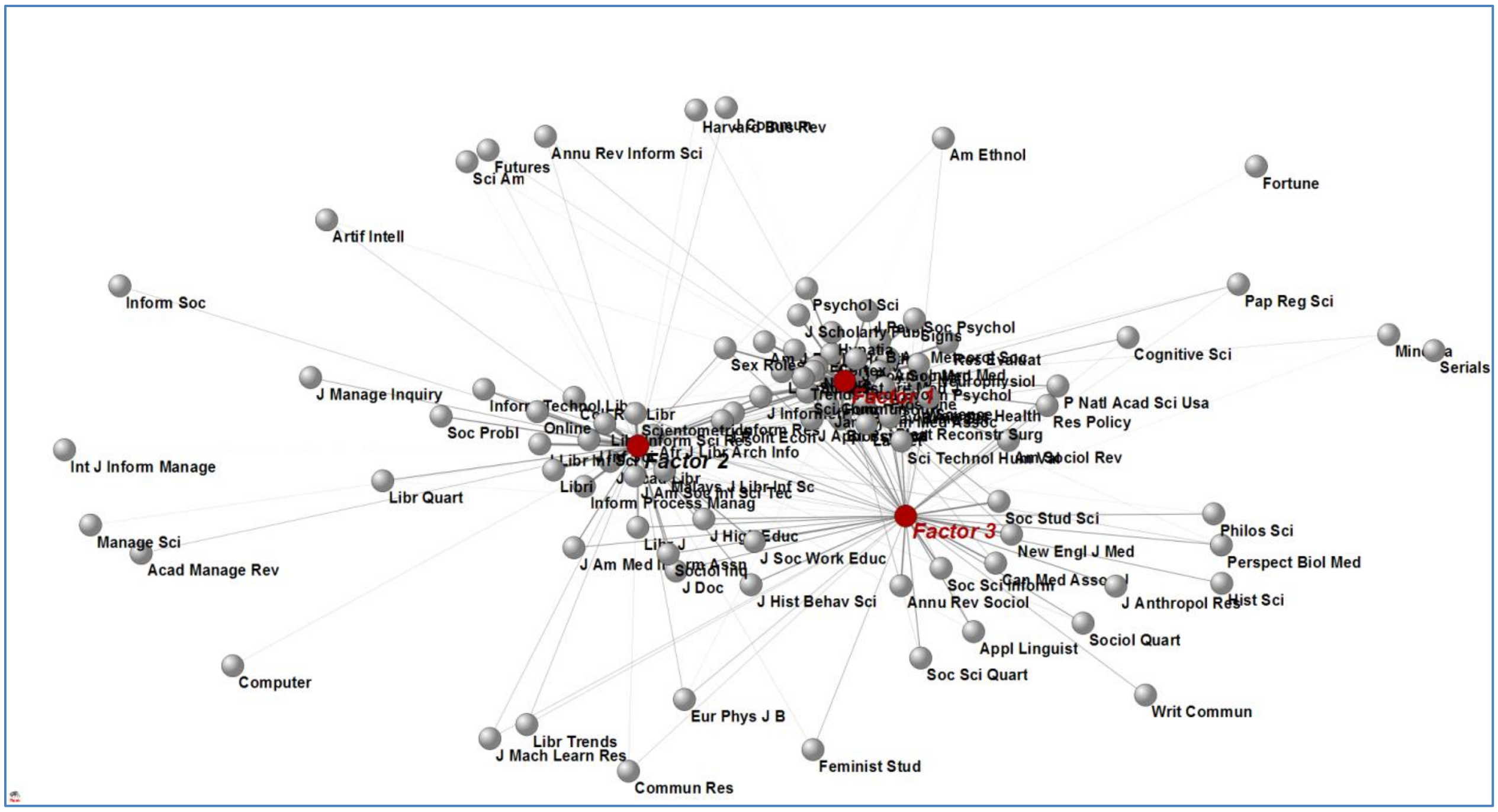

Figure 1.8. 96 of the 109 journal abbreviations with positive factor loadings.

## Cited References

Abramson, N. (1963). *Information Theory and Coding*. New York, etc.: McGraw-Hill.

Ahlgren, P., Jarneving, B., & Rousseau, R. (2003). Requirements for a Cocitation Similarity Measure, with Special Reference to Pearson's Correlation Coefficient. *Journal of the American Society for Information Science and Technology, 54*(6), 550-560.

Amsterdamska, O., & Leydesdorff, L. (1989). Citations: Indicators of Significance? *Scientometrics 15*(5-6), 449-471.

Ashby, W. R. (1958). Requisite variety and its implications for the control of complex systems. *Cybernetica*, *1*(2), 83-99.

Bar-Hillel, Y. (1955). An Examination of Information Theory. *Philosophy of Science*, *22*, 86-105.

Bateson, G. (1972). *Steps to an Ecology of Mind*. New York: Ballantine.

Bernstein, B. (1971). *Class, Codes and Control, Vol. 1: Theoretical studies in the sociology of language*. London: Routledge & Kegan Paul.

Blondel, V. D., Guillaume, J. L., Lambiotte, R., & Lefebvre, E. (2008). Fast unfolding of communities in large networks. *Journal of Statistical Mechanics: Theory and Experiment, 8*(10), 10008.

Borch, C. (2011). *Niklas Luhmann*. London and New York: Routledge.

Brillouin, L. (1962). *Science and Information Theory*. New York: Academic Press.

Brooks, D. R., & Wiley, E. O. (1986). *Evolution as Entropy*. Chicago/London: University of Chicago Press.

Burt, R. S. (1982). *Toward a Structural Theory of Action*. New York, etc.: Academic Press.

Callon, M. (1998). An essay on framing and overflowing: economic externalities revisited by sociology *The Laws of the Market* (pp. 244-269.). Oxford and Malden, MA: Blackwell.

Callon, M., Courtial, J.-P., Turner, W. A., & Bauin, S. (1983). From Translations to Problematic Networks: An Introduction to Co-word Analysis. *Social Science Information 22*(2), 191-235.

Callon, M., Law, J., & Rip, A. (Eds.). (1986). *Mapping the Dynamics of Science and Technology*. London: Macmillan.

Campbell, D. T. (1960). Blind variation and selective retentions in creative thought as in other knowledge processes. *Psychological review, 67*(6), 380.

Cozzens, S. E. (1989). What do citations count? The rhetoric-first model. *Scientometrics, 15*(5), 437-447.

Cronin, B. (1981). The need for a theory of citing. *Journal of Documentation, 37*(1), 16-24.

Cronin, B. (1984). *The citation process. The role and significance of citations in scientific communication*. London: Taylor Graham.

Cronin, B. (1998). Metatheorizing citation. *Scientometrics, 43*(1), 45-55.

Deacon, T. W. (1997). *The Symbolic Species: The co-evolution of language and the brain*. New York/London: W.W. Norton & Company.

Dretske, F. I. (1981). *Knowledge and the flow of information*. Cambridge MA: MIT Press.

Fiske, J. (2011). *Introduction to communication studies (3rd ed.)*. New York: Routledge.

Garfield, E. (1979). *Citation Indexing: Its Theory and Application in Science, Technology, and Humanities*. New York: John Wiley.

Garner, W. R., & McGill, W. J. (1956). The relation between information and variance analyses. *Psychometrika, 21*(3), 219-228.

Giddens, A. (1979). *Central Problems in Social Theory*. London, etc.: Macmillan.

Habermas, J. (1987). Excursus on Luhmann's Appropriation of the Philosophy of the Subject through Systems Theory *The Philosophical Discourse of Modernity: Twelve Lectures* (pp. 368-385). Cambridge, MA: MIT Press.

Hayles, N. K. (1990). *Chaos Bound; Orderly Disorder in Contemporary Literature and Science* Ithaca, etc.: Cornell University.

Hesse, M. (1980). *Revolutions and Reconstructions in the Philosophy of Science*. London: Harvester Press.

Hummon, N. P., & Doreian, P. (1989). Connectivity in a citation network: The development of DNA theory. *Social Networks, 11*(1), 39-63.

Husserl, E. ([1935/36] 1962). *Die Krisis der Europäischen Wissenschaften und die Transzendentale Phänomenologie*. Den Haag: Martinus Nijhoff.

Ivanova, I. A., & Leydesdorff, L. (2014). Redundancy Generation in University-Industry-Government Relations: The Triple Helix Modeled, Measured, and Simulated. *Scientometrics, 99*(3), 927-948.

Jakulin, A. (2005). *Machine learning based on attribute interactions* (Vol. http://stat.columbia.edu/~jakulin/Int/jakulin05phd.pdf). Ljubljana: University of Ljubljana.

Kamada, T., & Kawai, S. (1989). An algorithm for drawing general undirected graphs. *Information Processing Letters*, *31*(1), 7-15.

Kaplan, N. (1965). The norms of citation behavior: Prolegomena to the footnote. *American Documentation, 16*(3), 179-184.

Krippendorff, K. (1986). *Information Theory. Structural Models for Qualitative Data*. Beverly Hills, etc.: Sage.

Krippendorff, K. (2009a). W. Ross Ashby's information theory: a bit of history, some solutions to problems, and what we face today. *International Journal of General Systems, 38*(2), 189-212.

Krippendorff, K. (2009b). Information of Interactions in Complex Systems. *International Journal of General Systems, 38*(6), 669-680.

Lakatos, I., & Musgrave, A. (Eds.). (1970). *Criticism and the Growth of Knowledge* Cambridge: Cambridge University Press).

Law, J., & Lodge, P. (1984). *Science for Social Scientists*. London, etc.: Macmillan.

Leydesdorff, L. (1997). Why Words and Co-Words Cannot Map the Development of the Sciences. *Journal of the American Society for Information Science, 48*(5), 418-427.

Leydesdorff, L. (1998). Theories of Citation? *Scientometrics, 43*(1), 5-25.

Leydesdorff, L. (2006). The Biological Metaphor of a (Second-order) Observer and the Sociological Discourse. *Kybernetes, 35*(3/4), 531-546.

Leydesdorff, L. (2010a). Luhmann Reconsidered: Steps towards an empirical research program in the sociology of communication. In C. Grant (Ed.), *Beyond Universal Pragmatics: Essays in the Philosophy of Communication* (pp. 149-173). Oxford: Peter Lang.

Leydesdorff, L. (2010b). Redundancy in Systems which Entertain a Model of Themselves: Interaction Information and the Self-organization of Anticipation. *Entropy, 12*(1), 63-79.

Leydesdorff, L. (2011a). "Structuration" by Intellectual Organization: The Configuration of Knowledge in Relations among Scientific Texts. *Scientometrics 88*(2), 499-520.

Leydesdorff, L. (2011b). "Meaning" as a sociological concept: A review of the modeling, mapping, and simulation of the communication of knowledge and meaning. *Social Science Information, 50*(3-4), 1-23.

Leydesdorff, L. (2012). Radical Constructivism and Radical Constructedness: Luhmann's Sociology of Semantics, Organizations, and Self-Organization. *Constructivist Foundations, 8*(1), 85-92.

Leydesdorff, L. (2014a). Science Visualization and Discursive Knowledge. In B. Cronin & C. Sugimoto (Eds.), *Beyond Bibliometrics: Harnessing Multidimensional Indicators of Scholarly Impact* Cambridge MA: MIT Press.

Leydesdorff, L. (2014b). Niklas Luhmann's Magnificent Contribution to the Sociological Tradition: The Emergence of the Knowledge-Based Economy as an Order of Expectations. In T. Bakken & M. Tzaneva (Eds.), *Nachtflug der Eule: 150 Stimmen zum Werk von Niklas Luhmann*. Berlin: Lidi Verlag.

Leydesdorff, L., & Ahrweiler, P. (2014). In search of a network theory of innovations: Relations, positions, and perspectives. *Journal of the Association for Information Science and Technology, 65*(11), 2359-2374.

Leydesdorff, L., & Amsterdamska, O. (1990). Dimensions of Citation Analysis. *Science, Technology & Human Values, 15*(3), 305-335.

Leydesdorff, L., & Goldstone, R. L. (2014). Interdisciplinarity at the Journal and Specialty Level: The changing knowledge bases of the journal *Cognitive Science*. *Journal of the Association for Information Science and Technology 65*(1), 164-177.

Leydesdorff, L., & Ivanova, I. A. (2014). Mutual Redundancies in Inter-human Communication Systems: Steps Towards a Calculus of Processing Meaning. *Journal of the Association for Information Science and Technology, 65*(2), 386-399.

Leydesdorff, L., & Zawdie, G. (2010). The Triple Helix Perspective of Innovation Systems. *Technology Analysis & Strategic Management, 22*(7), 789-804.

Luhmann, N. (1975). Interaktion, Organisation, Gesellschaft: Anwendungen der Systemtheorie. In M. Gerhardt (Ed.), *Die Zukunft der Philosophie* (pp. 85-107). München: List.

Luhmann, N. (1990). Meaning as Sociology's Basic Concept. In N. Luhmann (Ed.), *Essays on Self-Reference* (pp. 21-79). New York / Oxford: Columbia University Press.

Luhmann, N. (1995). *Social Systems*. Stanford, CA: Stanford University Press.

Luhmann, N. (2002). How Can the Mind Participate in Communication? In W. Rasch (Ed.), *Theories of Distinction: Redescribing the Descriptions of Modernity* (pp. 169–184). Stanford, CA: Stanford University Press.

Luhmann, N. (2012). *Theory of Society, Vol. 1*. Stanford, CA: Stanford University Press.

Luukkonen, T. (1997). Why has Latour's theory of citations been ignored by the bibliometric community? Discussion of sociological interpretations of citation analysis. *Scientometrics, 38*(1), 27-37.

MacKay, D. M. (1969). *Information, Mechanism and Meaning*. Cambridge and London: MIT Press.

McGill, W. J. (1954). Multivariate information transmission. *Psychometrika, 19*(2), 97-116.

Nicolaisen, J. (2007). Citation analysis. *Annual review of information science and technology, 41*(1), 609-641.

Nöth, W. (2014). Human Communication from the Semiotic Perspective. In T. Dousa & F. Ibekwe-SanJuan (Eds.), *Theories of Information, Communication and Knowledge* (pp. 97-119). Dordrecht: Springer.

Otte, E., & Rousseau, R. (2002). Social network analysis: a powerful strategy, also for the information sciences. *Journal of Information Science, 28*(6), 441-453.

Padgett, J. F., & Powell, W. W. (2012). The Problem of Emergence *The Emergence of Organizations and Markets* (pp. 1-32). Princeton, NJ: Princeton University Press.

Parsons, T. (1951). *The Social System*. New York: The Free Press.

Parsons, T. (1968). Interaction: I. Social Interaction. In D. L. Sills (Ed.), *The International Encyclopedia of the Social Sciences* (Vol. 7, pp. 429-441). New York: McGraw-Hill.

Parsons, T., & Shils, E. A. (1951). *Toward a General Theory of Action*. New York: Harper and Row.

Popper, K. R. ([1935] 1959). *The Logic of Scientific Discovery*. London: Hutchinson.

Popper, K. R. (1972). *Objective Knowledge. An Evolutionary Approach*. Oxford: Oxford University Press.

Rip, A. (1981). A cognitive approach to science policy. *Research Policy, 10*(4), 294-311.

Shannon, C. E. (1948). A Mathematical Theory of Communication. *Bell System Technical Journal, 27*, 379-423 and 623-656.

Shannon, C. E., & Weaver, W. (1949). *The Mathematical Theory of Communication*. Urbana: University of Illinois Press.

Simon, H. A. (1973). The Organization of Complex Systems. In H. H. Pattee (Ed.), *Hierarchy Theory: The Challenge of Complex Systems* (pp. 1-27). New York: George Braziller Inc.

Strand, Ø., & Leydesdorff, L. (2013). Where is Synergy in the Norwegian Innovation System Indicated? Triple Helix Relations among Technology, Organization, and Geography. *Technological Forecasting and Social Change, 80*(3), 471-484.

Strydom, P. (1999). Triple Contingency: The theoretical problem of the public in communication societies. *Philosophy & Social Criticism, 25*(2), 1-25.

Theil, H. (1972). *Statistical Decomposition Analysis*. Amsterdam/ London: North-Holland.

Ulanowicz, R. E. (1986). *Growth and Development: Ecosystems Phenomenology*. San Jose, etc.: toExcel.

Ulanowicz, R. E. (2009). The dual nature of ecosystem dynamics. *Ecological modelling, 220*(16), 1886-1892.

van den Daele, W., & Weingart, P. (1975). Resistenz und Rezeptivität der Wissenschaft—zu den Entstehungsbedingungen neuer Disziplinen durch wissenschaftliche und politische Steuerung. *Zeitschrift für Soziologie, 4*(2), 146-164.

Vlieger, E., & Leydesdorff, L. (2011). Content Analysis and the Measurement of Meaning: The Visualization of Frames in Collections of Messages. *The Public Journal of Semiotics, 3*(1), 28.

Von Foerster, H. (2003). On Self-Organizing Systems and Their Environments *Understanding Understanding: Essays on Cybernetics and Cognition* (pp. 1-19). New York: Springer.

Weaver, W. (1949). Some Recent Contributions to the Mathematical Theory of Communication. In C. E. Shannon & W. Weaver (Eds.), *The Mathematical Theory of Communication* (pp. 93-117.). Urbana: University of Illinois Press.

Weinstein, F., & Platt, G. M. (1969). *The Wish to be Free: Society, Psyche, and Value Change*. Berkeley: University of California Press.

Whitley, R. D. (1984). *The Intellectual and Social Organization of the Sciences*. Oxford: Oxford University Press.

Woolgar, S. (1991). Beyond the citation debate: towards a sociology of measurement technologies and their use in science policy. *Science and Public Policy 18,* 319-326.

Wouters, P. (1998). The signs of science. *Scientometrics, 41*(1), 225-241.

Wouters, P. (1999). *The Citation Culture*. Amsterdam: Unpublished Ph.D. Thesis, University of Amsterdam.

Yeung, R. W. (2008). *Information Theory and Network Coding*. New York, NY: Springer.